\pdfoutput=1

\documentclass[aps,pre,reprint,superscriptaddress]{revtex4-1}

\usepackage[caption=false]{subfig}
\usepackage{graphicx}
\usepackage{amsmath}
\usepackage[hidelinks]{hyperref}

\newcommand*\diff{\mathop{}\!\mathrm{d}}

\DeclareMathOperator{\Tr}{Tr}

\begin{document}

\title{Dynamical Landau-de Gennes Theory for Electrically-Responsive Liquid Crystal Networks}

\author{Guido L. A. Kusters}
\email[]{g.l.a.kusters@tue.nl}
\affiliation{Department of Applied Physics, Eindhoven University of Technology, The Netherlands}

\author{Inge P. Verheul}
\affiliation{Department of Mathematics and Computer Science, Eindhoven University of Technology, The Netherlands}

\author{Nicholas B. Tito}
\affiliation{Electric Ant Lab, Amsterdam, The Netherlands}

\author{Paul van der Schoot}
\affiliation{Department of Applied Physics, Eindhoven University of Technology, The Netherlands}

\author{Cornelis Storm}
\affiliation{Department of Applied Physics, Eindhoven University of Technology, The Netherlands}
\affiliation{Institute for Complex Molecular Systems, Eindhoven University of Technology, The Netherlands}

\date{\today}

\begin{abstract}
Liquid crystal networks combine the orientational order of liquid crystals with the elastic properties of polymer networks, leading to a vast application potential in the field of responsive coatings, e.g., for haptic feedback, self-cleaning surfaces and static and dynamic pattern formation. Recent experimental work has further paved the way toward such applications by realizing the fast and reversible surface modulation of a liquid crystal network coating upon in-plane actuation with an AC electric field \cite{liu2017protruding,van2019morphing}. Here, we construct a Landau-type theory for electrically-responsive liquid crystal networks and perform Molecular Dynamics simulations to explain the findings of these experiments and inform on rational design strategies. Qualitatively, the theory agrees with our simulations and reproduces the salient experimental features. We also provide a set of testable predictions: the aspect ratio of the nematogens, their initial orientational order when cross-linked into the polymer network and the cross-linking fraction of the network all increase the plasticization time required for the film to macroscopically deform. We demonstrate that the dynamic response to oscillating electric fields is characterized by two resonances, which can likewise be influenced by varying these parameters, providing an experimental handle to fine-tune device design.
\end{abstract}

\pacs{aaa}

\maketitle

\section{Introduction}
Liquid crystal networks (LCNs), such as nematic elastomers, combine the orientational properties of liquid crystals with the elastic properties of polymer networks by incorporating liquid crystalline mesogens into the cross-linked polymer matrix \cite{finkelmann1981investigations}. The resulting coupling between the strain imposed on the material and the orientation of the liquid crystalline mesogens it contains introduces novel features that cannot entirely be traced back to either of the LCN constituents \cite{warner1996nematic}. Striking examples include large spontaneous deformations upon a temperature-induced phase transition of the liquid crystalline mesogens \cite{tajbakhsh2001spontaneous} and soft elasticity, i.e., macroscopic deformation at (almost) no energetic cost by reorientation of the liquid crystalline mesogens \cite{zentel1986shape}.

The theoretical description of these materials was pioneered by Warner and coworkers, who proposed a framework combining classical elasticity theory with the Landau-de Gennes theory for liquid crystals \cite{warner1988theory,warner1991elasticity}. This was later expanded upon to illustrate, among other things, the occurrence of phase transitions and instabilities in LCNs \cite{bladon1993transitions,verwey1996elastic,tajbakhsh2001spontaneous}, why these materials display soft deformation modes \cite{warner1994soft,olmsted1994rotational,verwey1995soft}, and how they respond to electric fields \cite{terentjev1994orientation}. Warner and Terentjev provide a comprehensive overview of the most important results in their monograph \cite{warner2007liquid}.

A deeper understanding of LCNs has led to the rapid development toward industrial applications of these materials in (soft) actuators \cite{ikeda2007photomechanics,white2015programmable}, surgical interventions \cite{gao2015biocompatible,zeng2017self,saed2017high} and 3D printing shape-memory liquid crystal elastomers \cite{kotikian20183d}. Most of these exploit the susceptibility of the mesogens to external stimuli such as heat or electric fields \cite{prost1995physics}. Recently, LCNs were also recognised as a prime candidate for stimuli-responsive coatings due to their tunable dynamical behavior, with envisaged applications in the fields of haptic feedback, self-cleaning surfaces and finely controlled pixel-like deformations (so-called voxels) \cite{smith2008review,heo2015fast}. Desirable properties for these applications are a swift response and large susceptibility to the applied stimulus. 

Considerable work has already been done to develop photoresponsive coatings \cite{white2012light,gelebart2018photoresponsive}, containing azobenzene moieties that undergo a \textit{trans}-to-\textit{cis} interconversion under illumination of appropriate wavelengths, as well as thermoresponsive coatings based on the isotropic-nematic phase transition of the liquid crystal network \cite{wermter2001liquid,li2006artificial,babakhanova2018liquid}. Although such coatings provide great versatility due to the possibility of remote and dynamical actuation, the range of pre-programmed topographical profiles and the time scale of deformation remain limited. Recent work by Liu \textit{et al.} proposes an electrically-responsive LCN coating in an attempt to mitigate these limitations \cite{liu2017protruding}, further paving the way toward dynamical, finely-resolved control of topographic profiles. In their work, Liu \textit{et al.} employ a transparent LCN film, superimposed on an array of interdigitated comb electrodes. The LCN consists, in addition to the polymer component, in large part of two species of liquid-crystal mesogen, namely (i) relatively immobile mesogens incorporated into the polymer network as permanent crosslinks and (ii) end-on grafted side-group mesogens. The latter of these have attached to them a cyano group resulting in a strong permanent dipole moment, whereas the former do not carry a permanent dipole moment. Both mesogen species are present in approximately equal amounts, and are prepared with homeotropic alignment in the film. Upon \textit{in-plane} actuation with an AC electric field in between the electrodes, perpendicular to the director field of the mesogens, the authors report (i) the fast and reversible formation of surface corrugations pre-programmed by the electrode placement, (ii) a surface height modulation of up to several percent in magnitude and (iii) a clear dependence of the modulation magnitude on the driving frequency. Thus, based on purely in-plane stimuli an out-of-plane expansion of the LCN is observed. Follow-up experiments by Van der Kooij \textit{et al.} on the same experimental system have since shown that there is a characteristic plasticization time associated with the initial macroscopic deformation of the LCN \cite{van2019morphing}.

Although one might expect electrothermal or electrostrictive effects to play an important role here, it can be shown that neither can adequately account for the magnitude of the observed modulation \cite{liu2017protruding,van2019morphing,krakovsky1999few,yamwong2002electrostrictive}. Instead, Liu \textit{et al.} rationalise the observed behavior by remarking that, upon in-plane actuation by the electric field, the end-grafted liquid crystalline mesogens reorient slightly away from their as-prepared homeotropic alignment, increasing their mutual excluded volume with the mesogens that cannot reorient as easily due their incorporation in the background polymer network. This increases the volume unavailable to the centers of mass of the mesogens. Indeed, if the two elongated mesogens are oriented perpendicularly to each other, they effectively occupy more volume than if their orientations are parallel, as was already argued by Lars Onsager in the  1940s \cite{onsager1949effects}. 

The physical picture that emerges, is one of molecular voids opening up following mesogen reorientation, an overall volume increase and subsequent viscoelastic relaxation of that volume increase as a function of time by the filling in of the voids by non-mesogenic (polymeric) material. These molecular voids are often referred to as ``free volume", i.e., the total system volume minus the total hard-core volume of the molecules comprising the film \cite{liu2017protruding,white2008high,liu2012light,liu2015new,van2019morphing}. The result of switching on an electric field perpendicular to the director field of the mesogens is a transient macroscopic, \textit{transverse} expansion of the film by virtue of the high density of mesogenic component in the LCN. Interestingly, an oscillating electric field produces the largest expansion, which arguably ties in with suppressing the viscoelastic relaxation of the host polymer matrix.

Although the coarse-grained Molecular Dynamics (MD) simulations of Liu \textit{et al.} \cite{liu2017protruding}, which probe the orientational properties of the LCN on the molecular scale, make it plausible that excluded volume indeed governs LCN expansion \cite{liu2017protruding}, a comprehensive theoretical framework remains lacking. In particular, such a framework may prove invaluable in informing on rational design strategies, by identifying experimentally relevant parameters dictating the effectiveness of the LCN modulation, quantifying the qualitative understanding we currently have based on simulations, and minimising costly trial and error in experiments. This last point is especially pressing, given the fact that research on the proposed set-up is ongoing and topical \cite{van2019morphing}.

For this purpose we construct a time-dependent Landau-de Gennes-type theory for electrically-responsive LCNs. After noting the failure of the conventional (neo-)classical theory of nematic elastomers to capture the experimental phenomenology, we instead take the connection between excluded volume and the LCN expansion as the basis for our description. We subsequently compare the resulting static and dynamic response with results from MD simulations building on earlier work of one of us \cite{liu2017protruding}, which show qualitative agreement. We demonstrate that the dynamic response to oscillating electric fields is characterized by two resonances, and that the time scales governing the LCN actuation can be influenced by varying experimental parameters, such as the aspect ratio of the mesogens, the cross-linking fraction of the network, and the degree of initial orientational order cross-linked into the liquid crystal network. Finally, we link back to the work of Liu \textit{et al.} \cite{liu2017protruding}, and detail how our theory explains some of their major findings.

The remainder of this paper is structured as follows. In section \ref{sec:model}, we construct an equilibrium model for our LCN and carry out MD simulations for comparison, focusing on the static response of the LCN. We illustrate the qualitative agreement between the two. Subsequently, we extend our model to a time-dependent description in section \ref{sec:dynamics}, and qualitatively validate the resulting dynamic response to turning on a DC electric field by comparing with our simulations. We find that there is a characteristic plasticization time associated with the macroscopic deformation of the LCN, in accordance with experimental findings \cite{van2019morphing}, which we argue to depend on the aspect ratio of the mesogens, the cross-linking fraction of the network, and the degree of initial orientational order cross-linked into the liquid crystal network. Next, in section \ref{sec:AC actuation}, we demonstrate that the dynamic response to oscillating electric fields is characterized by two resonances, corresponding to two distinct modes of mesogen reorientation, and we find that the associated time scales vary with the experimental parameters in the same manner as the plasticization time does. We use this qualitative picture to explain how the expansion of the LCN varies with the electric field strength and driving frequency in the simulations and in the experiments carried out by Liu \textit{et al.} \cite{liu2017protruding}. Finally, in section \ref{sec:conclusion}, we summarize our most important findings, and provide suggestions for simulations and experiments to further pave the way toward industrial applications.

\section{Model ingredients\label{sec:model}}
We describe our LCN by means of a Landau-de Gennes-type theory, rooted in the symmetry of the underlying liquid crystal. This provides a convenient avenue for constructing a qualitative theory, without the need for in-depth knowledge of the microscopic properties of the system at hand. Although the obvious approach to modelling the LCN would be to subsequently connect to classical elasticity theory, following Warner and Terentjev \cite{warner2007liquid}, it turns out that such a model fails to reproduce the experimental findings. Indeed, experiments show a marked volume expansion, whereas the classical theory of nematic elastomers predicts approximately volume preserving deformations. We refer to Ref. \cite{Thesis} for further details. This suggests a different mechanism lies at the root of the experimental findings, motivating us to construct a minimal model description based on the interplay between excluded volume generation and the LCN expansion.

As we are interested in explicitly modeling the volume increase of the LCN upon actuation with an electric field, the proper statistical ensemble for our theory is the isothermal-isobaric ensemble. Although one would then expect the relevant thermodynamic potential to be the Gibbs free energy density, this cannot hold true in our case because we must treat the relative volume change of the LCN as a proper order parameter in our Landau-de Gennes-type theory to be able to extract useful information regarding its expansion. Accordingly, the proper thermodynamic potential must be a minimum with respect to this relative volume change, which we denote $\eta\equiv\left(V-V_0\right)/V_0$, with $V$ the system volume and $V_0$ the initial volume prior to any deformation, in thermodynamic equilibrium. With this stipulation in mind, the relevant thermodynamic potential becomes the Gibbs free energy per unit \textit{reference} volume, which we denote $g$ \cite{Gibbsexplain}. In what follows we do not write the usual pressure-volume contribution to this free energy density, as we are solely interested in the case of the absence of any excess pressure $p$, implying $p=0$. 

\begin{figure}[htbp]
	\centerline{
		\includegraphics[width=9cm]{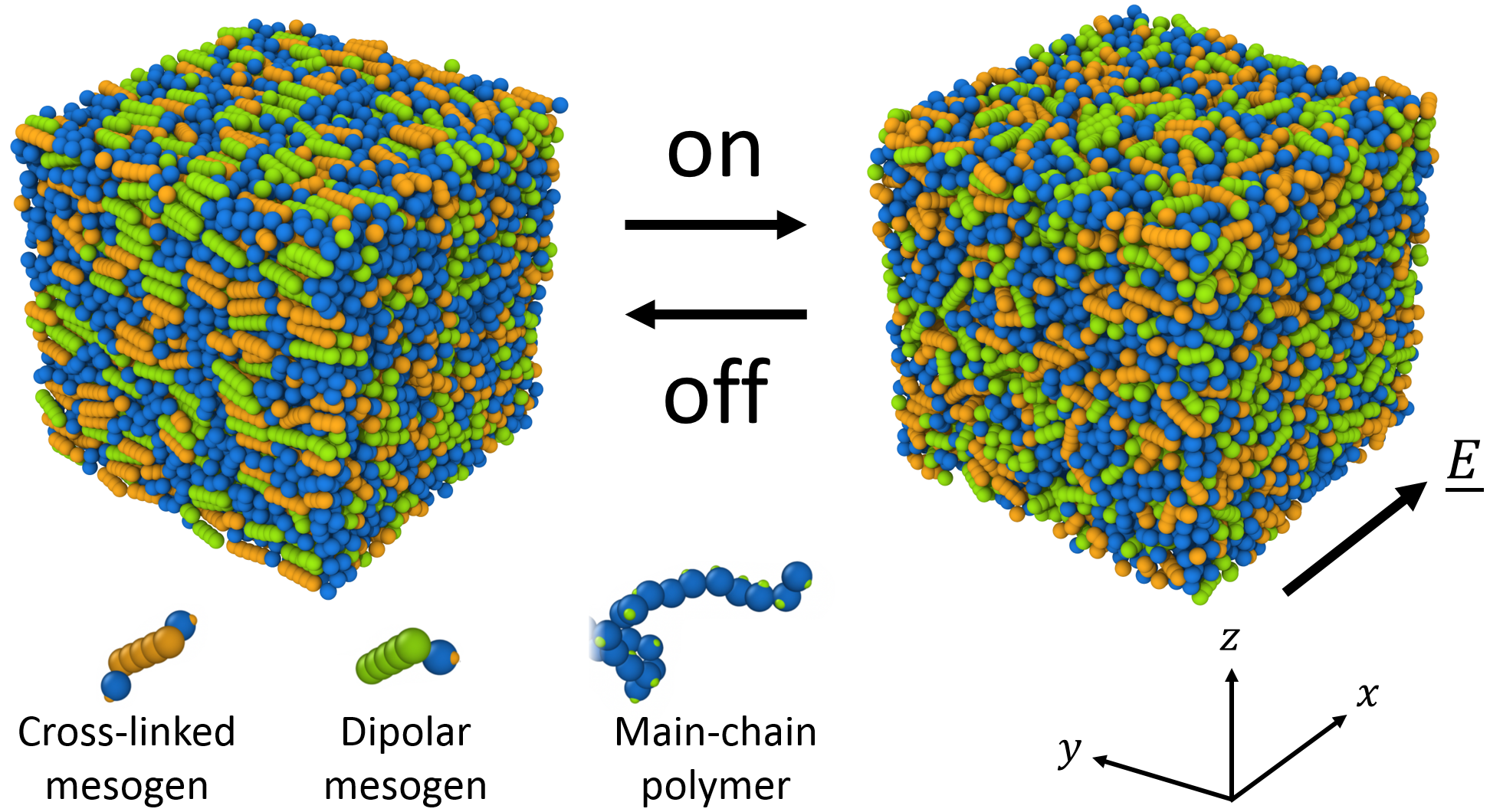}
	}
	\caption{Simulation snapshots visualising the problem geometry. Cross-linked mesogens are indicated in orange, dipolar mesogens in green and the main-chain polymer in blue. The colored dots drawn on the beads indicate binding sites, which are used as adhesion points during \textit{in-situ} polymerization. The mesogens are prepared with the director field along the $y$-axis and reorient upon actuation with an electric field along the $x$-direction, increasing the LCN volume.}
	\label{fig:diagram geometry}
\end{figure}

The material for which we aim to construct this thermodynamic potential consists of three key components, being the polymer component and two distinct species of mesogen schematically shown in Figure \ref{fig:diagram geometry} with the polymer component indicated in blue. The focus of the model is explicitly on the two different mesogen species, while we largely neglect the viscoelastic properties of the polymer network that experiments have shown result in the relaxation of changes in the LCN volume \cite{liu2017protruding,van2019morphing}; we return to this below when we investigate the dynamical behavior of the model. 

One of the mesogen species incorporated in the LCN is crosslinked into the polymer network on both ends, as indicated in orange in Figure \ref{fig:diagram geometry}, meaning such mesogens are immobile and can because of that not significantly reorient in response in an externally applied electric field. The other mesogen species is grafted end on onto the polymer network, as indicated in green in Figure \ref{fig:diagram geometry}, resulting in much more mobile dangling side groups. This species has on top of that attached to it a permanent dipole moment, ensuring that it interacts more strongly with electric fields than the cross-linked mesogens do. This suggests that the cross-linked mesogens cannot respond strongly to any applied external electric field, justifying our assumption that only the dipolar mesogens effectively interact with electric fields. The interaction free energy of the dangling mesogens is quadratic in the electric field $\underline{E}$ for nematic liquid crystals, by virtue of the inversion symmetry of the nematic director. This proportionality applies even for a collection of mesogens with a permanent dipole moment, which formally breaks this inversion symmetry, provided that the electric field is sufficiently weak, and the nematic symmetry is not broken by polarisation of the mesogens \cite{Hexplain}. 

In addition to the different response of the two types of mesogen to electric fields, the model must also reflect the different ways in which the two mesogen species are connected to the polymer network. In particular, if the cross-linked mesogens reorient, for example due to excluded-volume interactions with neighboring dipolar mesogens, they are subject to an elastic restoring force from the polymer network. In contrast, reorientation of the dipolar mesogens is impeded much less significantly by the polymer network, because these mesogens are only connected to the network by one end. To capture this asymmetry we assume that the dipolar mesogens can reorient freely in the polymer network, whereas the cross-linked mesogens are subject to a harmonic spring potential centered about the orientations that were cross-linked into the LCN during its preparation in the absence of an electric field. Here, we refer specifically to the degree of orientational order to which the cross-linked mesogens have relaxed under the experimental conditions \cite{warner2007liquid}.

The presence of two distinct mesogen species already suggests we need, at the least, three order parameters in our theory to model the volume expansion of the LCN resulting from any changes in the orientations of the mesogens in response to the electric field. That is, in addition to the aforementioned order parameter $\eta$ to keep track of the relative volume change of the LCN, we also require two order parameters to inform on the orientational properties of the cross-linked and dipolar mesogens, respectively. The most general way to express such orientational order is by means of the symmetric and traceless tensor nematic order parameter $Q_{i,j}=\langle u_iu_j-\delta_{i,j}/3\rangle$, with $\underline{u}$ the mesogen orientation unit vector and $\delta_{i,j}$ the Dirac delta. This suggests the introduction of two tensor order parameters $Q_{i,j}^\text{(c)}$ and $Q_{i,j}^\text{(d)}$, corresponding to the cross-linked and dipolar mesogens, respectively, to describe the orientational order of the different mesogen species. 

In what follows we assume that the cross-linked mesogens, due to their inability to strongly respond to any applied electric field, remain approximately uniaxial even if we apply the electric field perpendicular to the director; below we illustrate that our MD simulations justify this approximation. We then recover the uniaxial tensor order parameter $Q_{i,j}^\text{(C)}=S_c\left(\delta_{i,y}\delta_{j,y}-\delta_{i,j}/3\right)$ for the cross-linked mesogens, where the director lies along the $y$ axis and $S_c\equiv\frac{3}{2}\langle\cos^2\theta\rangle-\frac{1}{2}$ denotes the scalar nematic order parameter, with $\theta$ the angle a test mesogen makes with the director field and $\langle\dots\rangle$ an angular average. It is clear, however, that the dipolar mesogens produce, in general, a biaxial configuration if we apply the electric field perpendicular to the director. Although the usual approach would be to then write down the corresponding tensor nematic order parameter in the biaxial form $Q_{i,j}^\text{(d)}=S_d\left(\delta_{i,y}\delta_{j,y}-\delta_{i,j}/3\right)+P/3\left(\delta_{i,x}\delta_{j,x}-\delta_{i,z}\delta_{j,z}\right)$, where the electric field along the $x$ axis produces a degree of biaxiality $P=\frac{3}{2}\langle\sin^2\theta\cos 2\phi\rangle$, with $\phi$ the angle a test mesogen makes with the electric field in the $x$-$z$ plane, our MD simulations motivate us to take a different approach here.

From our MD simulations we learn that upon application of an electric field perpendicular to the director field in which both mesogen species are prepared, a clear division arises between dipolar mesogens that align with the director field and dipolar mesogens that align with the electric field. This is presumably due to the presence of the cross-linked mesogens, which remain ordered predominantly along the director field, justifying our approximation of uniaxial order for this mesogen species. The excluded-volume interactions with these mesogens must then be overcome by the dipolar mesogens to align with the electric field. This finding motivates us to, instead of describing the biaxial order of the dipolar mesogens collectively, separately describe the uniaxial order of the collection of director-field-aligned dipolar mesogens and the collection of electric-field-aligned dipolar mesogens. 

We model this by associating the collection of nematic-director-field-aligned mesogens with a fraction $1-x^2$ of the dipolar mesogens and the collection of electric-field-aligned dipolar mesogens with the remaining fraction $x^2$. Treating $x^2$ as an order parameter in our model then allows us to interpolate between these different populations of dipolar mesogen, in effect signifying a bi-stable equilibrium between the two. Here, the use of the quadratic form $x^2\geq 0$ ensures this fraction is always non-negative; the constraint $x^2\leq1$ we enforce by hand. We remark that, of course, taken together these populations of dipolar mesogens again exhibit biaxial order; the main advantage of treating this biaxial order as a bi-stable equilibrium is the ability to also capture the interpolation between the two populations, i.e., the reorientation of the dipolar mesogens, which our MD indicate to be important.

\begin{figure}[htbp]
	\centerline{
		\includegraphics[width=9cm]{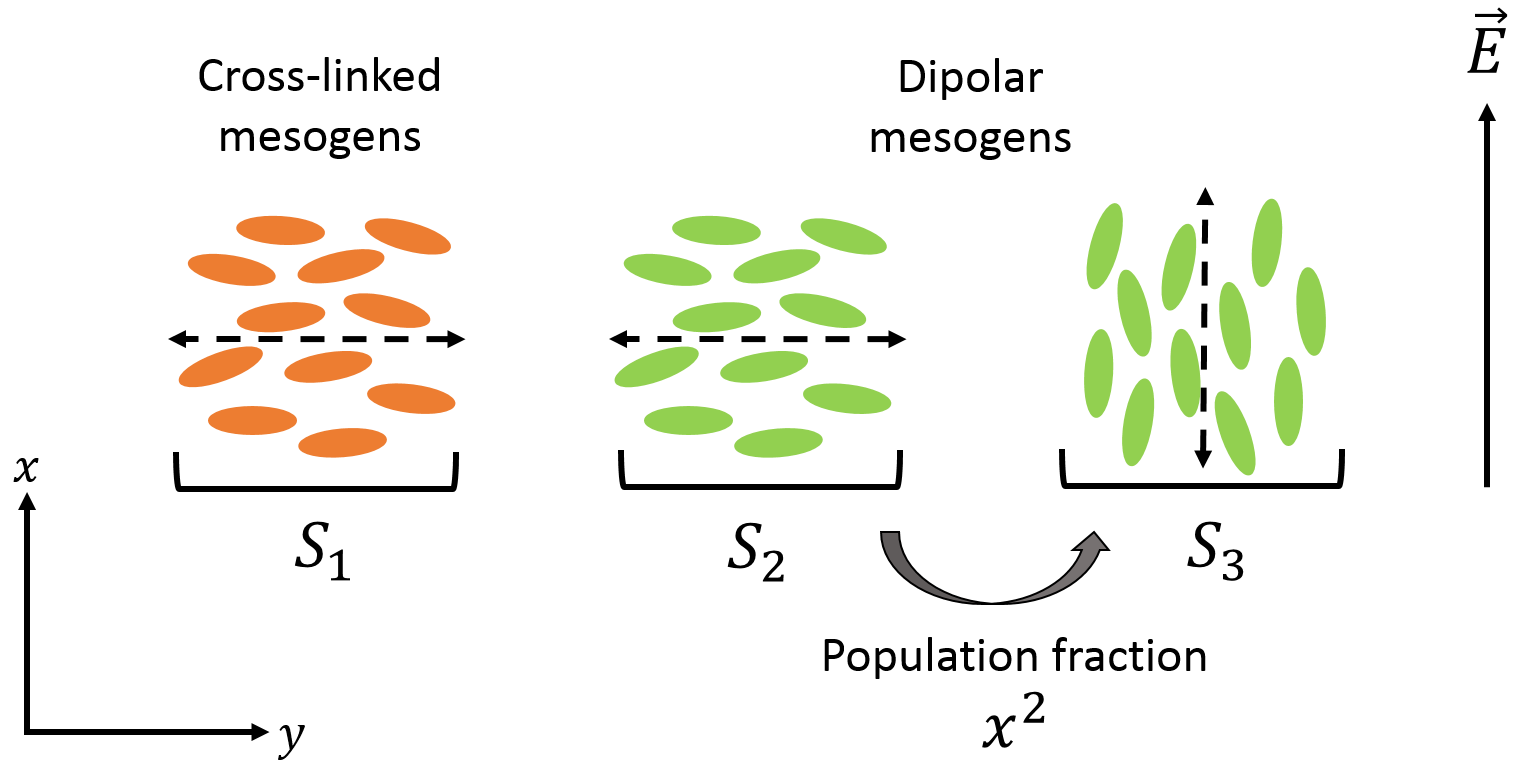}
	}
	\caption{Schematic representation of the model mesogens. Cross-linked mesogens are indicated with orange ellipsoids and dipolar mesogens are indicated with green ellipsoids. The scalar nematic order parameters of the cross-linked mesogens, $S_1$, and the director-field-aligned dipolar mesogens, $S_2$, are measured along the $y$-axis (dashed arrows), along which both mesogen species are prepared. The scalar nematic order parameter of the electric-field-aligned dipolar mesogens, $S_3$, is measured along the $x$-axis (dashed arrow), along which the electric field is applied. The population fraction $x^2$ denotes the fraction of dipolar mesogens that aligns with the electric field.}
	\label{fig:schematic model}
\end{figure}

The above suggests that we need not use two tensor nematic order parameters, $Q_{i,j}^\text{(c)}$ and $Q_{i,j}^\text{(d)}$, the latter of which is in general biaxial, to describe the orientational order of the cross-linked and dipolar mesogens, respectively. Instead, we use three \textit{uniaxial} tensor nematic order parameters for the orientational order of the cross-linked mesogens, the collection of director-field-aligned dipolar mesogens and the collection of electric-field-aligned dipolar mesogens. Relabeling for ease of reference, we denote these $Q_{i,j}^\text{(1)}$, $Q_{i,j}^\text{(2)}$ and $Q_{i,j}^\text{(3)}$, respectively. The former two have their director along the $y$ axis, along which both mesogen species are prepared, whereas the director of the latter lies along the $x$ axis, along which the electric field is applied. We denote the corresponding scalar nematic order parameters, as measured along their respective directors, $S_1$, $S_2$ and $S_3$, which are schematically shown in Figure \ref{fig:schematic model}.

Although this implies that we can capture the orientational order of the cross-linked and dipolar mesogens as $Q_{i,j}^\text{(c)}=Q_{i,j}^\text{(1)}$ and $Q_{i,j}^\text{(d)}=\left(1-x^2\right)Q_{i,j}^\text{(2)}+x^2Q_{i,j}^\text{(3)}$, respectively, and the global orientational order as $Q_{i,j}=\left(Q_{i,j}^\text{(c)}+Q_{i,j}^\text{(d)}\right)/2$ on account of there being equal amounts of both in the network, we will not construct our Landau theory in terms of the invariants of these tensors. The reason for this is that since these tensor nematic order parameters are defined along different axes, their explicit combination turns out to introduce unphysical couplings. In particular, coupling terms promoting cross-hatched, i.e., perpendicular, orientational order of the different mesogen species and populations arise in the nematic phase, in which we are interested \cite{Thesis}. The alternative we follow here is to construct our model in terms of the invariants of the individual tensor nematic order parameters of the different mesogen species and populations we distinguish, i.e., $Q_{i,j}^\text{(1)}$, $Q_{i,j}^\text{(2)}$ and $Q_{i,j}^\text{(3)}$ for the cross-linked mesogens, the director-field-aligned dipolar mesogens and the electric-field-aligned dipolar mesogens. A similar division of different types of nematogens into populations was in fact already used by Onsager as early as 1949 in his seminal work \cite{onsager1949effects}. 

Clearly, the invariants that comprise our model must also give rise to interactions between the different mesogen species and order parameters. Broadly speaking, these interactions come in two flavors: mesogen-mesogen interactions and mesogen-volume interactions. The first of these couple the orientational properties of the cross-linked mesogens to those of the dipolar mesogens via excluded-volume-like interactions. This has the effect of promoting the alignment between the different mesogen species, as their mutual excluded volume is smallest when they align. Assuming identical rod-like shapes for both mesogens with length $L$ and diameter $D$, the mutual excluded volume of two mesogens is proportional to $L^2D+\mathcal{O}\left(LD^2\right)$, we expect the same dependence on the dimensions of the mesogens for this interaction \cite{onsager1949effects}. A similar argument can be made if we increase the cross-linking fraction of the network, while keeping the total amount of mesogens constant, i.e., if we increase the fraction of mesogens that are fully cross-linked into the polymer network at the expense of dipolar mesogens. Then, any dipolar mesogen will have an increased number of immobile cross-linked mesogens as its neighbors. This makes it increasingly difficult for the electric field to overcome the excluded-volume interactions of the dipolar mesogen with its neighbors in order to induce reorientation. Thus, the excluded-volume mesogen-mesogen interactions must be stronger in this case, at least regarding the reorientation of the dipolar mesogens. We remark that the reverse effect of any cross-linked mesogen interacting with a decreasing number of dipolar mesogens is much less significant. The reason for this is that these mesogens are fully cross-linked into the polymer network, and cannot respond strongly to the applied electric field.

The second type of interaction between order parameters we incorporate, couples the orientational properties of the mesogens to the volume increase of the LCN. This interaction is closely related to the excluded-volume interactions discussed above, as in sufficiently dense mesogenic systems the excluded volume generated upon mesogen reorientation contributes to the total volume of the system. To reflect this, we associate configurations in which the different mesogen species become increasingly aligned with a volume decrease, whereas the volume increases if the mesogens become orientationally disordered.

Taken together, the above ingredients define a Gibbs free energy per unit reference volume for our model as a function of the following order parameters: (i) the relative volume change of the LCN, $\eta$, (ii) the tensor nematic order parameters $Q_{i,j}^\text{(1)}$, $Q_{i,j}^\text{(2)}$ and $Q_{i,j}^\text{(3)}$ for the cross-linked mesogens, the director-field-aligned dipolar mesogens and the electric-field-aligned dipolar mesogens, and (iii) the population fraction $x^2$. Note that since the tensor nematic order parameters are all presumed to be uniaxial, all information on the magnitude of orientational order is encapsulated in the associated scalar nematic order parameters $S_1$, $S_2$ and $S_3$, in terms of which we express our model below.

\section{Theory\label{sec:theory}}
Taking all the contributions described in the section above into account, our model Gibbs free energy per unit reference volume becomes
\begin{equation}\label{eq:gpop}
    g=\sum_{i=1}^{12} g_i,
\end{equation}
where we will discuss the form of each contribution $g_i$ in turn. The first three contributions correspond simply to the free energy associated with each mesogen species or population, according to
\begin{equation}\label{eq:g1-3}
\begin{split}
    g_1&=g_\text{LdG}\left(S_1\right),\\
    g_2&=\left(1-x^2\right)g_\text{LdG}\left(S_2\right),\\
    g_3&=x^2g_\text{LdG}\left(S_3\right).
\end{split}
\end{equation}
Here, 
\begin{equation}\label{eq:standardLT}
    g_\text{LdG}(S_i)=\frac{1}{2}AS_i^2-\frac{1}{3}BS_i^3+\frac{1}{4}CS_i^4, \, \, \, \, \, \, \, \, \, \, i=1,2,3
\end{equation}
denotes the standard Landau-de Gennes free energy density for uniaxial nematic liquid crystals, with $A,B,C$ phenomenological coefficients, as follows from a power series expansion in terms of $\Tr\underline{\underline{Q}}^{\text{(i)}^n}$, where $\Tr$ denotes the trace operator and the series is truncated after $n=4$ \cite{prost1995physics,gramsbergen1986landau}. 

Although the phenomenological coefficient $A$ is generally taken to depend explicitly on temperature, to reflect the temperature-induced isotropic-nematic phase transition, we are currently not interested in modeling this dependence. In fact, prior simulations and experiments suggest the temperature-dependent behavior of the LCN is significantly more complicated than can be expected from a simple Landau theory \cite{liu2017protruding,van2019morphing}; our own MD simulations confirm this. Instead, we choose $A$ such that the configuration in which the LCN is prepared, given by $S_1=S_2=S_0$, $x^2=0$ and $\eta=0$ minimizes the total free energy \eqref{eq:gpop} in the absence of an electric field. Here, $S_0$ denotes the as-prepared scalar nematic order parameter, the value of which we specify further below by comparing with the simulations, and the value of $S_3$ matters not as there are no electric-field-aligned dipolar mesogens in the absence of an electric field. As will become apparent below, this statement is identical to choosing $A$ such that $S_0$ minimizes equation \eqref{eq:standardLT}, since the remaining contributions to the free energy are automatically minimized under these conditions. This effectively eliminates $A$ from our model description in favor of $S_0$.

Finally, note that in equation \eqref{eq:g1-3} the free-energy contribution of equation \eqref{eq:standardLT} is multiplied by factors of $\left(1-x^2\right)$ and $x^2$ for the free energy associated with the director-field-aligned dipolar mesogens $g_2$ and the electric-field-aligned dipolar mesogens $g_3$, ensuring only the fraction of dipolar mesogens that is in a given population contributes to its free energy. As we presume there to be equal amounts of the cross-linked and dipolar mesogens, reflecting the experimental work of Liu \textit{et al.} \cite{liu2017protruding}, their free energies bear equal weight.

The next contribution encodes the interaction of the dipolar mesogens with the electric field, which, on symmetry grounds, must be proportional to the product of the electric field $\underline{E}$ with the electric susceptibility tensor $\underline{\underline{\Sigma}}$ of said mesogens, according to $-\underline{E} \, \underline{\underline{\Sigma}} \, \underline{E}$. Since the electric susceptibility of the dipolar mesogens is closely related to their orientational order, this contribution takes the form \cite{prost1995physics,gramsbergen1986landau}
\begin{equation}\label{eq:g4}
    g_4=-x^2HS_3,
\end{equation}
where $H\propto\lvert\underline{E}\rvert^2$ acts only on the electric-field-aligned dipolar mesogens, not the director-field-aligned dipolar mesogens. This is because the electric field is applied perpendicular to the director field, and so the associated contribution to the free energy is small due to the low susceptibility of the mesogens along this axis. For this reason $H$ actually represents a relative value, rather than an absolute one. This approximation is also in line with the philosophy of the model, in which we have two populations of dipolar mesogens, which order along perpendicular axes. Indeed, a similar coupling between $S_2$ and $H$ would promote orientational order of the director-field-aligned dipolar mesogens perpendicular to, rather than along, their prescribed director axis.

The difference in how the different mesogen species interact with their surroundings is underscored by the elastic restoring force from the polymer network the cross-linked mesogens are subject to. The corresponding contribution penalizes any deviation from the orientational properties the cross-linked mesogens were prepared with in the absence of an electric field, given by $\underline{\underline{Q}}^\text{(1)}=\underline{\underline{Q}}^\text{(0)}$, with $\underline{\underline{Q}}^\text{(0)}$ the uniaxial tensor nematic order parameter corresponding to the as-prepared configuration of the LCN. The lowest-order tensor invariant that achieves this effect is of the form $\Tr\left(\underline{\underline{Q}}^\text{(1)}-\underline{\underline{Q}}^\text{(0)}\right)^2$, resulting in the expression
\begin{equation}\label{eq:g5}
    g_5=\frac{1}{2}\kappa\left(S_1-S_0\right)^2,
\end{equation}
which acts as a harmonic spring, with $\kappa$ a phenomenological spring constant, exercising a restoring force that pulls the orientational order parameter of the cross-linked mesogens $S_1$ back toward the orientational order $S_0$ they had, when relaxed under the experimental conditions prior to the application of an electric field. Although technically not identical, as noted before, $S_0$ can be thought of in similar terms as the orientational order of the cross-linked mesogens when they were cross-linked into the polymer network. 

We follow a similar procedure in writing down the excluded-volume-like interactions between the different mesogen species, but now substituting $\underline{\underline{Q}}^\text{(0)}$ with $\underline{\underline{Q}}^\text{(2)}$ and $\underline{\underline{Q}}^\text{(3)}$, to yield
\begin{equation}\label{eq:g6-7}
    \begin{split}
        g_6&=\frac{1}{2}\left(1-x^2\right)\lambda\left(S_1-S_2\right)^2,\\
        g_7&=\frac{1}{2}x^2\lambda\left(S_1^2+S_1S_3+S_3^2\right),
    \end{split}
\end{equation}
where the former indicates interactions between cross-linked mesogens and director-field-aligned dipolar mesogens, whereas the latter concerns interactions between cross-linked mesogens and electric-field-aligned dipolar mesogens. Here, $\lambda$ is a phenomenological parameter, which, as argued before, we expect to scale with the mesogen dimensions as $L^2D$, and to increase upon increasing the cross-linking fraction of the network. 

The first term in equation \eqref{eq:g6-7} is reminiscent of equation \eqref{eq:g5}, acting as a harmonic spring penalising misalignment between the different mesogen species. The second, however, looks markedly different, although it follows from exactly the same procedure and achieves the same effect. The reason for this is that the director for the electric-field-aligned dipolar mesogens lies along the electric field, along the $x$ axis, rather than the director field, along the $y$ axis, yielding $\Tr\left(\underline{\underline{Q}}^\text{(1)}-\underline{\underline{Q}}^\text{(3)}\right)^2\propto\left(S_1^2+S_1S_3+S_3^2\right)$ instead. 

We point out that we do not explicitly incorporate excluded-volume-like interactions between the different populations of dipolar mesogen because this would, again, promote orientational order of these populations perpendicular to their assigned director, undermining the philosophy of the model. Instead, the excluded-volume-like interactions between the different populations of dipolar mesogen are mediated by the population fraction $x^2$, such that if the orientational mismatch between the two becomes sufficiently stringent, the population fraction adjusts to mitigate this. This feature is already included qualitatively in equation \eqref{eq:g6-7}. The sceptical reader may take comfort in the fact that including a cross term between the different populations of dipolar mesogens does not significantly alter the qualitative behavior of the model, provided that the associated coupling constant remains moderate (results not shown).

As already announced, the interaction between the orientational order of the mesogens and the relative volume increase of the LCN is closely related to the excluded volume of the mesogens in our model. However, instead of penalising misalignment between the mesogens, we must now penalise the volume of the LCN when the mesogens become increasingly aligned. Thus, instead of constructing the invariants of the difference between tensor order parameters we now take their sums $\Tr\left(\underline{\underline{Q}}^\text{(1)}+\underline{\underline{Q}}^\text{(2)}\right)^2$ and $\Tr\left(\underline{\underline{Q}}^\text{(1)}+\underline{\underline{Q}}^\text{(3)}\right)^2$, resulting in the contributions
\begin{equation}\label{eq:g8}
\begin{split}
    g_8&=\left(1-x^2\right)\xi\eta\left(S_1+S_2\right)^2,\\
    g_9&=x^2\xi\eta\left(S_1^2-S_1S_3+S_3^2\right),\\
    g_{10}&=-\xi\eta\left(4S_0^2\right).
\end{split}
\end{equation}
Here, we have multiplied the various contributions by the volume order parameter $\eta$ to ensure the proper effect on the system volume, $\xi$ is a phenomenological coefficient and we have again separated the contributions associated with the different populations of dipolar mesogen apparent from $g_8$ and $g_9$. We ignore explicit cross terms, as we did before, to ensure the model remains internally consistent. The final contribution shown in equation \eqref{eq:g8} ensures the orientational order of the mesogens is measured relative to the initial configuration by subtracting from the contributions $g_8$ and $g_9$ their counterparts with $S_1=S_2=S_0$ and $x^2=0$.

Finally, we must ensure that the free energy of equation \eqref{eq:gpop} is bounded from below, which leads us to add the additional terms
\begin{equation}\label{eq:g9-10}
    \begin{split}
        g_{11}&=\frac{1}{2}B_1 x^4,\\
        g_{12}&=\frac{1}{2}B_0\eta^2,
    \end{split}
\end{equation}
representing bulk-modulus-like terms in the population fraction $x^2$ and the relative volume increase $\eta$. Note that, technically, the former is not required to keep the free energy bounded from below since we constrain $0\leq x^2\leq1$. However, we do need it to allow $x^2$ to take intermediate values between $0$ and $1$ upon minimizing equation \eqref{eq:gpop} with respect to all order parameters.

It is clear that the model outlined above in principle requires for a very large parameter space to be explored, potentially giving rise to very rich behavior (see also below). However, the parameter space can be significantly reduced by scaling the theory to make it dimensionless. To this end, we introduce the set of scaled scalar nematic order parameters $s_i\equiv S_i/\left(B^2/2C\right)$, with $i=1,2,3$, and the scaled free volume order parameter $\tilde{\eta}\equiv\eta/\left(\xi B^2/B_0C^2\right)$, where we leave the population fraction $x^2$ unchanged; these order parameters are subject to the scaled electric field strength $h\equiv H/\left(B^3/27C^2\right)$. This procedure reduces the set of phenomenological parameters characterizing our model to the parameter combinations $\tilde{\kappa}\equiv\kappa/\left(B^2/C\right)$, associated with the the elastic coupling of the cross-linked mesogens to the polymer network, $\tilde{\lambda}\equiv\lambda/\left(B^2/C\right)$, associated with the excluded-volume interactions between the different mesogen species, $\zeta\equiv\xi^2/B_0C$, associated with the coupling between the orientational order of the mesogens and the LCN volume, $\tilde{B_1}\equiv B_1/\left(B^4/C^3\right)$, corresponding to the bulk-modulus free energy penalty associated with the population fraction $x^2$, and $s_0\equiv S_0/\left(B/2C\right)$, which represents the initial degree of orientational order, crosslinked into the network. Recall that we eliminate the phenomenological coefficient $A$ from the model description by demanding that the initial configuration, with $S_1=S_2=S_0$, $x^2=0$ and $\eta=0$, minimizes the free energy density per unit reference volume \eqref{eq:gpop} in the absence of an electric field, implying that $A=BS_0-CS_0^2$. The result of this scaling procedure is a \textit{universal} curve for the theory in terms of the scaled order parameters $s_1$, $s_2$, $s_3$, $x^2$ and $\tilde{\eta}$, dependent only on the set of scaled parameters $\tilde{\kappa}$, $\tilde{\lambda}$, $\zeta$, $\tilde{B_1}$ and $s_0$, as well as the scaled electric field strength $h$.

Although the parameter space available to the scaled theory is still significant, extensive numerical tests, carried out by numerically minimizing the scaled free energy per unit reference volume \eqref{eq:gpop} using the ``fsolve'' root-finding algorithm in the programming language Python, show that there is a broad range of parameter values that produce the key features of the results of both our MD simulations and the experiments (results not shown). We point out that in carrying out the minimization procedure, we fix the value of order parameters that exceed their permitted value range, i.e., $0\leq x^2\leq1$. Although a similar constraint applies to the scalar nematic order parameters $-1/2\leq S_i\leq1$, with $i=1,2,3$, we cannot explicitly enforce this constraint for their scaled counterparts $s_i$, since the effective values of the scaling coefficients $B$ and $C$ are unknown for the LCN under consideration. For the purpose of this paper we focus on parameter values representative of the regimes described above. 

Before we present the corresponding equations of state, we first pause to elucidate the simulation protocol for our MD simulations, as these will serve as a reference for the theory in the remainder of this paper. In our simulations, we work in terms of the fundamental Lennard-Jones units for distance $\mathcal{D}$, energy $\mathcal{E}$, mass $\mathcal{M}$ and time $\mathcal{T}=\sqrt{\mathcal{M}\mathcal{D}^2/\mathcal{E}}$. The electric field strength is expressed in terms of $\sqrt{\mathcal{E}/\left(4\pi\epsilon_0\mathcal{D}^3\right)}$, with $\epsilon_0$ the vacuum permittivity. We use the HOOMD-blue package (v2.1.1) to simulate the liquid crystal network within the isothermal-isobaric ($NpT$) ensemble \cite{anderson2008general,glaser2015strong}, using periodic boundary conditions. As visualised in Figure \ref{fig:diagram geometry}, we represent the polymer main-chain (blue) as strings of particles, held together by harmonic bonds
\begin{equation}
    U_\text{bond}\left(r\right)=\frac{1}{2}k\left(r-r_0\right)^2,
\end{equation}
with $k=1110 \, \mathcal{E}/\mathcal{D}^2$ and $r_0=0.9 \, \mathcal{D}$ \cite{peter2006thickness}, acting on their distance of separation $r$. In addition, we represent the cross-linked mesogens (orange) and the dipolar mesogens (green) as collections of five particles held together by stiff harmonic bonds with $k=5000\mathcal{E}/\mathcal{D}^2$ and $r_0=0.5 \, \mathcal{D}$, and kept in a rod-like shape by harmonic potentials
\begin{equation}
    U_\text{ang}\left(\theta\right)=\frac{1}{2}k_\theta\left(\theta-\pi\right)^2,
\end{equation}
with $k_\theta=200 \, \mathcal{E}/\text{rad}^2$, acting on the angle $\theta$ between three particles. Our simulations contain $70$ polymer main-chains, each $50$ particles long, $1166$ mesogens of both species, and we incorporate no explicit or implicit solvent. Finally, all particles interact with each other by means of Lennard-Jones interactions
\begin{equation}
    U_\text{LJ}\left(r\right)=4\epsilon\left[\left(\frac{r}{\sigma}\right)^{12}-\left(\frac{r}{\sigma}\right)^6\right]+U_{\text{LJ},0},
\end{equation}
with $\epsilon=1 \, \mathcal{E}$ and $\sigma=1 \, \mathcal{D}$, which are cut off at $r_c=2.5 \, \mathcal{D}$; $U_{\text{LJ},0}$ is chosen such that $U_\text{LJ}\left(r_c\right)=0$.

The polymerization of the LCN is carried out \textit{in situ}, by assigning binding sites to the polymer main-chain particles (shown as green dots in Figure \ref{fig:diagram geometry}), and assigning to the terminal particles of both mesogen species the binding partner (shown as orange dots in Figure \ref{fig:diagram geometry}). The cross-linked mesogens contain binding partners on either end, whereas the dipolar mesogens only contain a binding partner on one end. To enforce one-to-one binding of binding sites and binding partners, we introduce first a repulsive inverse power potential
\begin{equation}
    U_\text{pow}\left(r\right)=4\epsilon\left(\frac{\sigma}{r}\right)^{12}+U_{\text{pow},0},
\end{equation}
which is cut off at $r_c=2.5 \, \mathcal{D}$, and $U_{\text{pow},0}$ is chosen such that $U_\text{pow}\left(r_c\right)=0$. This potential acts on pairs of binding sites and pairs binding partners (both with $\sigma=0.9 \, \mathcal{D}$ and $\epsilon=1 \, \mathcal{E}$), as well as on pairs containing both a binding site and a binding partner and pairs of the Lennard-Jones particles themselves (both with $\sigma=0.6 \, \mathcal{D}$ and $\epsilon=1 \, \mathcal{E}$). We next add a strongly attractive Gaussian potential 
\begin{equation}
    U_\text{Gauss}\left(r\right)=\epsilon_\text{site}\exp\left[-\frac{1}{2}\left(\frac{r}{\sigma_\text{site}}\right)^2\right],
\end{equation}
with $\epsilon_\text{site}=-500 \, \mathcal{E}$, $\sigma_\text{site}=0.2 \, \mathcal{D}$, and cut-off length $r_c=3 \, \mathcal{D}$, acting only on pairs containing both a binding site and a binding partner. Finally, to achieve the desired degree of mesogen alignment during crosslinking, equal and opposite unity charges $q=1 \, \sqrt{4\pi\epsilon_0 \mathcal{E}\mathcal{D}}$ are added to the ends of \textit{both} the cross-linked and the dipolar mesogens, and a dummy electric field $E$ is applied along the $x$-axis. The electric field strength can be varied to change the degree of mesogen alignment and the charges serve only to couple to the electric field; no Coulombic interactions between the dipoles are taken into account. 

We then crosslink the system in the presence of the dummy electric field by repeatedly performing short simulations in the $NVE$ ensemble and permanently connecting overlapping binding sites and partners. This process is repeated until a desired fraction of the binding sites ($85\%$, in our case) has been permanently bound to a partner. Here, the stoichiometry is such that there are on average two bonds to a mesogen of either species for every three particles on the polymer main-chain, emulating the experimental system, dense in mesogenic component. Following this, all binding sites and partners, as well as the associated potentials $U_\text{pow}$ and $U_\text{Gauss}$, are removed from the simulation.

After crosslinking, the LCN is equilibrated in the $NpT$ ensemble in the presence of the dummy electric field, using the Martyna-Tobias-Klein barostat-thermostat \cite{martyna1994constant}, with temperature and pressure coupling constants set to unity. We use $T=0.35 \, \mathcal{E}/k_B$ for the temperature, $p=1 \, \mathcal{E}/\mathcal{D}^3$ for the pressure, $\diff t=0.001\mathcal{T}$ for the time step size, and we equilibrate for $10^4$ time steps. Subsequently, the charges are removed from the cross-linked mesogens, which are inert in our simulations, and we carry out a second $NpT$ equilibration for $10^4$ time steps in the absence of the dummy electric field. This yields the initial configuration of the LCN for the simulations presented in this paper, for which we use a total simulation time of $300\mathcal{T}$. We remark that we ensure that the simulation results are not particular to a specific topology by carrying out the procedure described above multiple times and each time introducing random distortions (permanent bonds) during the crosslinking process, resulting in different random network topologies for the LCN. For further details regarding the simulation protocol, we refer to Refs. \cite{liu2017protruding,BEPReport}.

This concludes a discussion of our theory. We now present our equations of state.

\section{Equations of state\label{sec:eos}}
\begin{figure}[htbp]
	\centerline{
		\includegraphics[width=9cm]{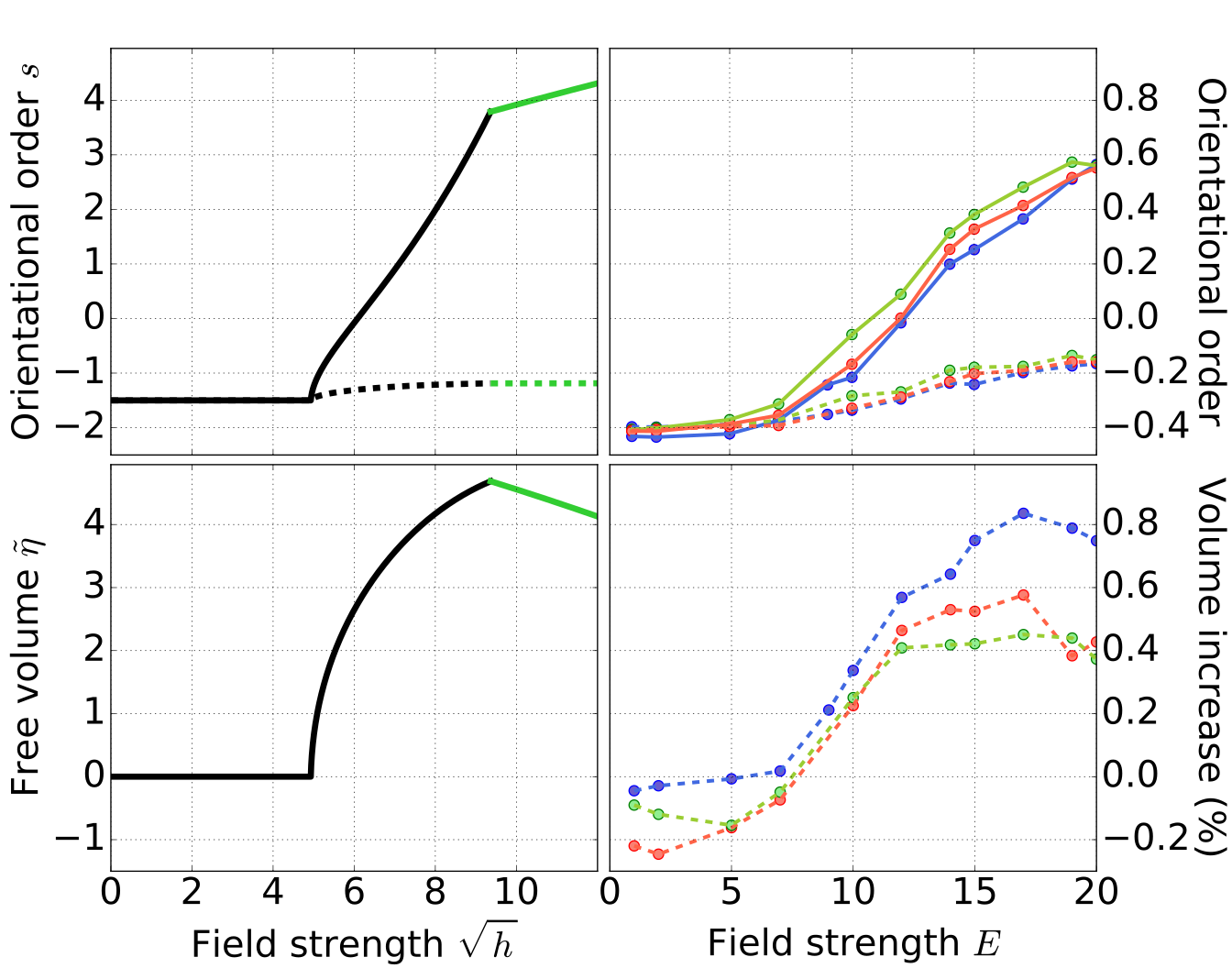}
	}
	\caption{Equations of state for the scaled theory (left) and MD simulations (right) as a function of electric field strength. The top panels show the orientational order parameter for the cross-linked (dashed lines) and dipolar mesogens (solid lines), measured along the electric field axis. The bottom panels show a measure for the concomitantly generated free volume. Theory: Black curves denote stable solutions and green curves indicate stable solutions with saturated population of electric-field-aligned dipolar mesogens $x^2=1$. We denote the scaled scalar nematic order parameter, measured along the electric field axis,  $s$, the scaled free volume order parameter $\Tilde{\eta}$ and the scaled electric field strength $\sqrt{h}\propto\lvert\underline{E}\rvert$. The explicit scalings for these, as well as for our model parameters, are given in the main text at the end of Section \ref{sec:theory}. The values we use for our scaled model parameters are $\Tilde{\kappa}=1.0$ for the elastic coupling of the cross-linked mesogens to polymer network, $\tilde{\lambda}=0.4$ for the excluded-volume-like coupling constant, $\zeta=0.03$ for the mesogen-volume coupling constant, $\tilde{B_1}=5.0$ for the bulk-modulus-like free energy penalty for the population fraction $x^2$ and $s_0=3.0$ for the initial degree of orientational order, crosslinked into the network. Simulations: the percentage volume increase is measured relative to the undeformed configuration prior to the application of the electric field. We work in terms of fundamental Lennard-Jones units, at fixed temperature $T=0.35$ and fixed pressure $p=1.0$, with time steps $\diff t=0.001$ and total simulation time $t_\text{sim}=300.0$; see the end of section \ref{sec:theory} for further details. Different colors denote different random network topologies and lines are guides to the eye.}
	\label{fig:phase diagram}
\end{figure}

Figure \ref{fig:phase diagram} shows the results of a numerical minimization of equation \eqref{eq:gpop}, scaled as described in the previous section, compared with the phase diagram we obtain from our MD simulations. The Figure illustrates good qualitative agreement between theory (left), where we show a single universal curve for the chosen parameter values, and simulations (right), where we show results for three randomly polymerized network topologies. Indeed, focusing first on the orientational properties of the LCN (top panels), we find for increasing electric field strength: (i) no to weak response to variations in the electric field strength for both mesogen species below a critical field strength, (ii) a critical field strength beyond which the response to variations in the electric field strength is strong, and (iii) saturation of the orientational order. This last regime is colored green in Figure \ref{fig:phase diagram} and corresponds to a saturated population of field-aligned dipolar mesogens, i.e., $x^2=1$, where we enforce the constraint $x^2\leq1$ by hand. Recall that a similar constraint for the scalar nematic order parameters $s_i$, with $i=1,2,3$, cannot be enforced because these have been scaled with coefficients of which the values are unknown.

The same qualitative features hold for the concomitantly generated free volume (bottom panels), albeit with a decrease in volume in the final regime rather than the saturation of the orientational order. This last feature, which gives rise to a maximum in the free volume curve, follows directly from the orientational order at the corresponding electric field strength: the maximum occurs at the point of maximum global orientational \textit{dis}order. We attribute the slight decrease in volume we find in our simulations at very weak electric field strengths (bottom-right panel) to the initial configuration in our simulations corresponding to a local minimum of the free energy, rather than the global free energy minimum. This is because relaxation of the LCN to thermodynamic equilibrium occurs on time scales slower than those accessible in our simulations; prior simulations and experiments also hint at this slow relaxation \cite{liu2017protruding,van2019morphing}. If we subsequently apply a weak electric field to this initial configuration, which is stable on the simulation time scales in the absence of an electric field, we then help the system to relax to a configuration with a lower free energy by decreasing its volume slightly. Noting that we do not aim to capture this last feature explicitly within the model, we find that the mechanism of free volume generation, on which the theoretical model is based, qualitatively explains simulation results and shows good qualitative agreement.

Although the static behavior of the model we show here serves as a proof of concept, supporting the role of excluded volume in driving the LCN to expand upon actuation, we still lack some key features of the simulations and the experiments. Most notably, the static model neglects the viscoelastic relaxation of the polymer network, which ensures the volume of the LCN relaxes as a function of time, if we apply a constant electric field. Accordingly, to achieve a steady-state volume increase of the LCN in the simulations and the experiments, alternating electric fields are used to continually actuate the LCN before it relaxes viscoelastically \cite{liu2017protruding,van2019morphing}. In what follows, we will be interested in how our model parameters, and in particular those that are experimentally accessible, influence the response of the LCN to alternating electric fields. To this end, we dedicate the next section to dynamics, where we first ensure our model dynamics properly reflects the viscoelastic relaxation of the LCN in response to the application of a constant electric field.

\section{Dynamics}\label{sec:dynamics}
The equilibrium model we describe in the previous sections can also be used to investigate the dynamical behavior of the LCN, by complementing it with a set of differential equations governing the temporal behavior of the order parameters. To this end, however, it is important to recall that the model, represented by equation \eqref{eq:gpop}, neglects the viscoelastic relaxation of the LCN, which is known to play an integral role in the volume expansion dynamics observed in the simulations and the experiments \cite{liu2017protruding,van2019morphing}. This viscoelastic relaxation is also the reason that AC actuation is needed to achieve steady-state LCN modulation in the simulations and the experiments, as the response following DC actuation simply relaxes. Thus, in order to connect with simulations and experiments, the dynamical equations characterizing the model must be adapted to account for this. For the non-conserved order parameters under consideration, the simplest form the associated differential equations can take describes relaxational dynamics \cite{hohenberg1977theory}, according to
\begin{equation}\label{eq:dynamicaleqs}
    \partial_\tau\psi=-\Gamma_\psi\left[\frac{\partial g}{\partial \psi}+\gamma\eta\langle\tau\rangle_\eta\delta_{\psi,\eta}\right]+\theta_\psi.
\end{equation}
Here, $\tau$ denotes time, $\psi$ represents the relevant order parameter, i.e., $S_1,S_2,S_3,x$ or $\eta$, $\Gamma_\psi$ denotes the associated kinetic coefficient, and $\theta_\psi$ represents a Gaussian noise term, which ensures that the fluctuation-dissipation theorem is satisfied \cite{glauber1963time}. The first term in brackets represents the mean-field dynamics following from the free energy density shown in equation \eqref{eq:gpop}, whereas the second term effectively corrects for the viscoelastic relaxation of the LCN, i.e., it ensures the free volume order parameter $\tilde{\eta}$ relaxes as a function of time. Here $\gamma$ is a phenomenological constant, $\langle\tau\rangle_\eta$ denotes the ensemble-averaged time the free volume has existed for, and $\delta_{\psi,\eta}$ the Kronecker delta.

Formally, the inclusion of this relaxational term implies that at thermodynamic equilibrium there \textit{cannot} exist any non-zero free volume. However, by virtue of the aforementioned slow relaxation of the LCNs under consideration \cite{liu2017protruding,van2019morphing}, we do not expect full relaxation to thermodynamic equilibrium to occur on the available time scales. Instead, in our MD simulations the temporal evolution of the system stops upon reaching some locally stable configuration. A similar construction for the scaled theory can be achieved by stopping the model dynamics after some time $\tau_\text{stop}$, shorter than the time scale required for full relaxation to thermodynamic equilibrium, has passed. This construction can be used to show that the qualitative features of the phase diagram in Figure \ref{fig:phase diagram} remain unchanged following the addition of free volume relaxation, provided that the system cannot fully relax to thermodynamic equilibrium on the used time scales \cite{Thesis}.

In addition, although the Gaussian noise term $\theta_\psi$ plays an integral part in the dynamics of some systems \cite{hamada1981dynamics,van1994noise,park1996noise,van1997nonequilibrium,lee2014critical}, here we limit ourselves to mean-field dynamics by setting $\theta_\psi=0$. This is justified, as the phase diagram we report on in Figure \ref{fig:phase diagram} does not explicitly show any free energy barriers, the absence of which we verified by scanning over the space of initial configurations. This indicates there are no kinetic traps, and that the LCN ends up in the proper equilibrium configuration regardless of noise. Thus, in what follows we suffice with explicitly solving only the deterministic part of equation \eqref{eq:dynamicaleqs}. We again non-dimensionalize the newly introduced (kinetic) parameters, and choose values within a broad regime $\Gamma_x<\Gamma_{S_1}<\Gamma_{S_2}=\Gamma_{S_3}<\Gamma_\eta$ exhibiting the key features of simulation and experimental findings. Here, we choose our scalings such that the scaled kinetic coefficient for the free volume $\Tilde{\Gamma}_{\Tilde{\eta}}=1$, i.e., we measure time relative to the dynamics of the scaled free volume order parameter $\tilde{\eta}$. We refer to the Appendix for the full scaling procedure.

\begin{figure}[htbp]
	\centerline{
		\includegraphics[width=9cm]{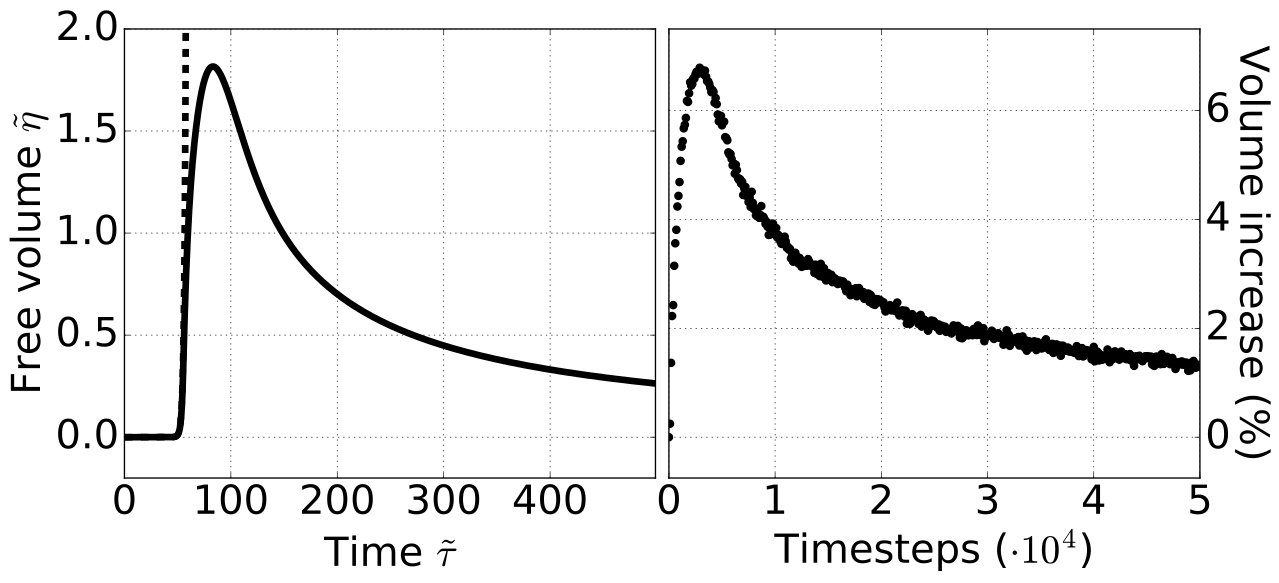}
	}
	\caption{Time traces of the generated free volume for the scaled theory (left) and MD simulations (right). The dashed line (left) indicates a short-time approximation obtained by linearizing the theory. Theory: we denote the scaled free volume order parameter $\Tilde{\eta}$, the scaled time $\Tilde{\tau}=\tau/\left[\xi^2/\Gamma_{\eta}B_0^2C\right]$ and we apply a scaled electric field strength $\sqrt{h}=18$. See the Appendix for the full scaling procedure for these, as well as for our model parameters. Parameter values used: $\Tilde{\kappa}=1.0$, $\tilde{\lambda}=0.4$, $\zeta=0.03$, $\tilde{B_1}=5.0$, $s_0=3.0$, $\delta x\left(0\right)^2=10^{-16}$ for the initial fraction of field-aligned dipolar mesogens, $\Tilde{\gamma}=0.001$ for the free volume relaxation parameter, and $\Tilde{\Gamma}_{s_1}=0.05$, $\Tilde{\Gamma}_{s_2}=\Tilde{\Gamma}_{s_3}=0.1$, $\Tilde{\Gamma}_{x}=0.01$ and $\Tilde{\Gamma}_{\Tilde{\eta}}=1.0$ for the scaled kinetic coefficients. Simulations: we work in terms of fundamental Lennard-Jones units, at fixed temperature $T=0.35$ and fixed pressure $p=1.0$, with time steps $\diff t=0.001$ and electric field strength $E=30.0$; see the end of section \ref{sec:theory} for further details.}
	\label{fig:time traces}
\end{figure}

Figure \ref{fig:time traces} shows the numerical solution of equation \eqref{eq:dynamicaleqs} for the scaled free volume order parameter $\tilde{\eta}$ if we turn on a DC electric field at $\tau=0$, compared with our MD simulations. This illustrates that the response to a DC electric field is remarkably similar for theory and simulations, both showing a sharp rise in free volume followed by its relaxation. A striking difference between the two, however, is the theoretical prediction of a pseudo lag time for free volume creation; this corresponds to the time required for a significant population of field-aligned dipolar mesogens to form and thus for stable actuation. We stress that this is by no means an \textit{actual} lag time, as exponential growth of the free volume order parameter $\tilde{\eta}$ starts as soon as the electric field is turned on; this is not visible in Figure \ref{fig:time traces} due to the linear scale. This finding already hints at the effective driving frequency range being limited by the intrinsic time scales of the LCN. Although this feature is absent from the MD simulations, in recent experimental work Van der Kooij \textit{et al.} report on a similar time scale that they associate with the plasticization of the polymer network \cite{van2019morphing}. In what follows we shall also refer to the pseudo lag time as such.

A possible explanation for the absence of the plasticization time from our MD simulations, which are small compared to the experimental system, is that fluctuations in our simulations may not be small. In this case the instability plasticizing the LCN is forced, such that we are unable to temporally resolve the existence of a plasticization time. The importance of such fluctuations in plasticizing the LCN is illustrated in the paragraphs below, where we more strongly connect our theoretical findings regarding this plasticization time with experiments.

To further substantiate the experimental relevance of the predicted plasticization time, and to illustrate the importance of fluctuations in plasticizing the LCN, we estimate the short-time relaxation dynamics by linearizing the dynamical equations \eqref{eq:dynamicaleqs} in the observables $S_1,S_2,S_3,x^2$ and $\eta$. That is, we expand about the initial configuration $S_1=S_0+\varepsilon \, \delta S_1$, $S_2=S_0+\varepsilon \, \delta S_2$, $S_3=S_0+\varepsilon \, \delta S_3$, $x^2=\varepsilon \, \left(\delta x\right)^2$ and $\eta=\varepsilon \, \delta\eta$, with $\varepsilon\ll1$, and solve the set of dynamical equations up to $\mathcal{O}\left(\varepsilon\right)$. The resulting exponential solution for the scaled free volume $\Tilde{\eta}$ is indicated by the dashed curve in Figure \ref{fig:time traces}. We find that the plasticization time roughly corresponds to the time required to saturate the field-aligned population of dipolar mesogens $x^2=1$ in the short-time approximation. The estimate then reads
\begin{equation}\label{eq:nucleation time}
    \tau_p\approx-\frac{1}{\Gamma_x S_0}\frac{1}{2H-3\lambda S_0}\log{\delta x\left(0\right)},
\end{equation}
where $\delta x\left(0\right)^2\ll1$ denotes the initial fraction of field-aligned dipolar mesogens. From the divergence of equation \eqref{eq:nucleation time} at $H=3\lambda S_0/2$ it is apparent that we recover a critical field strength, as was also the case for the phase diagram in Figure \ref{fig:phase diagram}. Above this critical field strength $\tau_p>0$, and so fluctuations in $x^2$ spontaneously grow under the influence of the electric field. Conversely, below the critical field strength $\tau_p<0$, indicating that fluctuations relax as a function of time.

Notably, equation \eqref{eq:nucleation time} explicitly shows the dependence of the plasticization time on experimentally accessible parameters. Although the dependence on the electric field strength $H$ is straightforward to interpret, by noting that it exerts a torque on the dipolar mesogens to induce reorientation, equation \eqref{eq:nucleation time} also suggests that decreasing the orientational order cross-linked into the network, $S_0$, generally decreases the plasticization time. Similarly, the dependence on the parameter $\lambda$, which scales with the magnitude of excluded volume interactions, suggests that a decrease in plasticization time can be achieved by decreasing the liquid crystalline mesogen aspect ratio or the cross-linking fraction.

Finally, equation \eqref{eq:nucleation time} clearly indicates that some initial disorder is important in driving free volume generation, as for a perfectly aligned initialization $\delta x\left(0\right)\rightarrow0$ the plasticization time diverges $\tau_p\rightarrow\infty$. Although in experiments and simulations the initial configuration is naturally characterised by small random fluctuations of the order parameters around the equilibrium values, in the theory we enforce this by setting $\delta x\left(0\right)>0$.

This establishes the dynamical response of our model LCN upon the application of a constant electric field. Below we extend this discussion to alternating electric fields, which are of most direct interest for future experiments and applications.

\section{AC actuation}\label{sec:AC actuation}
\begin{figure}[htbp]
	\centerline{
		\includegraphics[width=9cm]{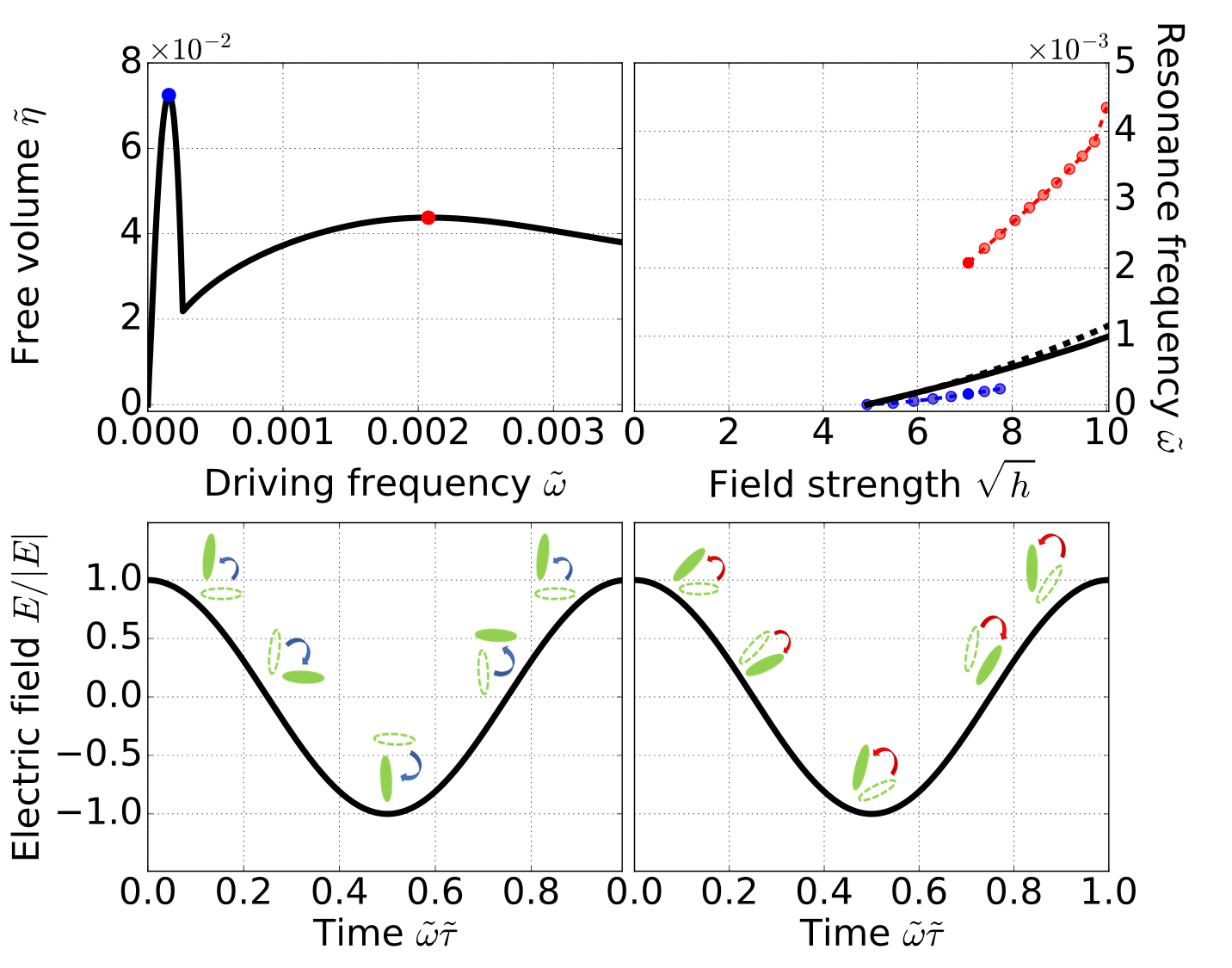}
	}
	\caption{Resonance frequencies for steady-state free volume gain as a function of electric field strength (top) and the corresponding microscopic mechanisms (bottom). Top left: steady-state gain in the scaled free volume order parameter $\Tilde{\eta}$ as a function of the scaled driving frequency $\Tilde{\omega}=\omega\left[\xi^2/\Gamma_{\eta}B_0^2C\right]$, for a scaled electric field strength $\sqrt{h}=7$; the colored dots denote resonance frequencies. Top right: colored dots show the scaled resonance frequencies $\Tilde{\omega}$ as a function of the scaled electric field $\sqrt{h}\propto\lvert\underline{E}\rvert$ in the regimes of low electric field strength (blue) and high electric field strength (red). The black curve denotes the numerically evaluated plasticization time at the same field strength (divided by 4 to correct for oscillation of the AC electric field), with the corresponding estimate equation \eqref{eq:nucleation time} indicated with the dashed line. See the Appendix for the full scaling procedure and see Figure \ref{fig:time traces} for the used parameter values. Bottom: schematic representation of the reorientation of the dipolar mesogens during a full oscillation of the electric field in the low-field regime (left, blue) and the high-field regime (right, red). The dipolar mesogens are indicated by green ellipsoids, with the dashed outlines denoting their previous orientation. The corresponding response of the free volume order parameter, which is also an important aspect of the resonances, is not shown. See the main text for an explanation.}
	\label{fig:actuation}
\end{figure}

The characteristic plasticization time is also relevant when applying an AC electric field $H\propto\left(\lvert\underline{E}\rvert\cos2\pi\omega\tau\right)^2$, with $\omega$ the driving frequency and $\tau$ the time, as illustrated by Figure \ref{fig:actuation}. The top-left panel shows, for a given electric field strength, how the steady-state free volume varies as a function of the driving frequency. This yields two temporally separated resonances for free volume generation: a low-frequency resonance (blue) and a high-frequency resonance (red).

The top-right panel in Figure \ref{fig:actuation} shows how these resonance frequencies vary as a function of the electric field strength (colored data points), as compared to the numerically evaluated plasticization time (solid black curve; the dashed line shows the estimate equation \eqref{eq:nucleation time}). We again observe no response below the critical field strength, followed by two distinct regimes: ``low" driving frequencies dominate at low field strengths (blue) and ``high" driving frequencies dominate at high field strengths (red). Note that we do not show the low-frequency resonance for high electric field strengths because it is completely dominated by the high-frequency resonance there, i.e., it is completely absorbed into the high-frequency resonance and no longer detectable. The same holds for the high-frequency resonance at low electric field strengths.

The existence of two resonances implies two coupled processes must underlie our findings: mesogen reorientation and the response of the volume to this reorientation. This claim is supported by the fact that the two distinct resonances persist even if we fix the nematic scalar order parameters to their as-prepared values $S_1=S_2=S_3=S_0$, indicating the fraction $x^2$ and the free volume order parameter $\eta$ dictate the most important dynamics. Although the non-linear nature of the model has not allowed us to analytically estimate the resonance frequencies in a meaningful way, a qualitative comparison of the resonance frequencies with the plasticization time yields additional insights. Indeed, despite the low-frequency resonance occurring on slightly slower time scales than the plasticization time, and the high-frequency resonance occurring on significantly faster time scales, both exhibit the qualitative trends apparent from equation \eqref{eq:nucleation time} upon variation of the model parameters. This provides an experimental handle to influence the resonant driving frequency.

We further probe the origin of these two resonances by investigating the temporal evolution of the fraction of field-aligned dipolar mesogens $x^2$ \cite{Thesis}, which indicates that in the low-frequency regime the dipolar mesogens follow the electric field. That is, for each cycle of the electric field, the dipolar mesogens completely reorient to align with the field when its magnitude is high, and fully relax when the field strength passes through zero. This behavior is combined with the dynamics of the free volume order parameter $\eta$, which describes how the volume dynamically ``wraps itself around" the space that is freed up by mesogen reorientation, to yield the resonance. The bottom-left panel of Figure \ref{fig:actuation} schematically illustrates the described low-frequency resonance dynamics, where we show only the response of the mesogen orientation to the electric field for visual clarity. This visualizes the low-frequency resonance dynamics of the order parameter $x^2$, but not that of the free volume order parameter $\eta$.

Conversely, the high-frequency regime corresponds to frequencies too fast for the dipolar mesogens to follow. Instead, upon turning on the electric field, the dipolar mesogens start to reorient to align with the field. However, before the reorientation is complete, the electric field has already decreased in magnitude significantly and passes through zero, and the dipolar mesogens relax briefly in response to this. Then, as the oscillation continues, the electric field strength increases again, and in response the dipolar mesogens reorient a bit further than on the previous cycle. This ``pumping" of the dipolar mesogens by the electric field continues until a steady state of reorientation and (partial) relaxation is achieved. Again, the resonance results from a combination of this behavior with the dynamics of the free volume order parameter $\eta$. The bottom-left panel of Figure \ref{fig:actuation} schematically illustrates the high-frequency resonance dynamics, where we again show only the response of the mesogen reorientation to the electric field, which visualizes the dynamics of the order parameter $x^2$, for visual clarity.

\begin{figure}[htbp]
    \subfloat{\includegraphics[width=4.cm]{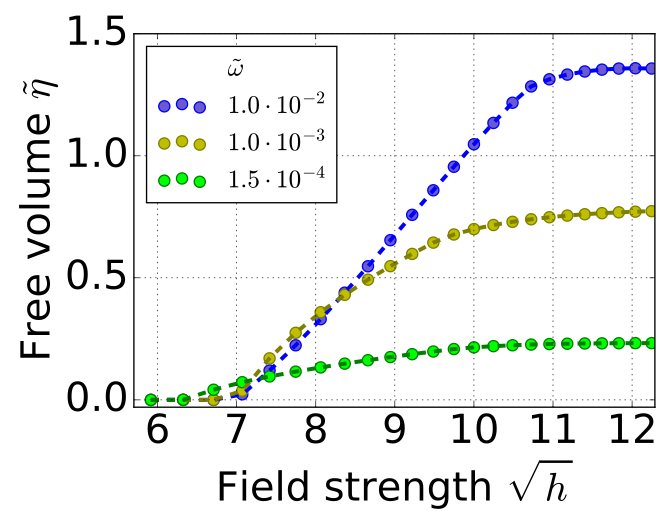}}
    \subfloat{\includegraphics[width= 4.cm]{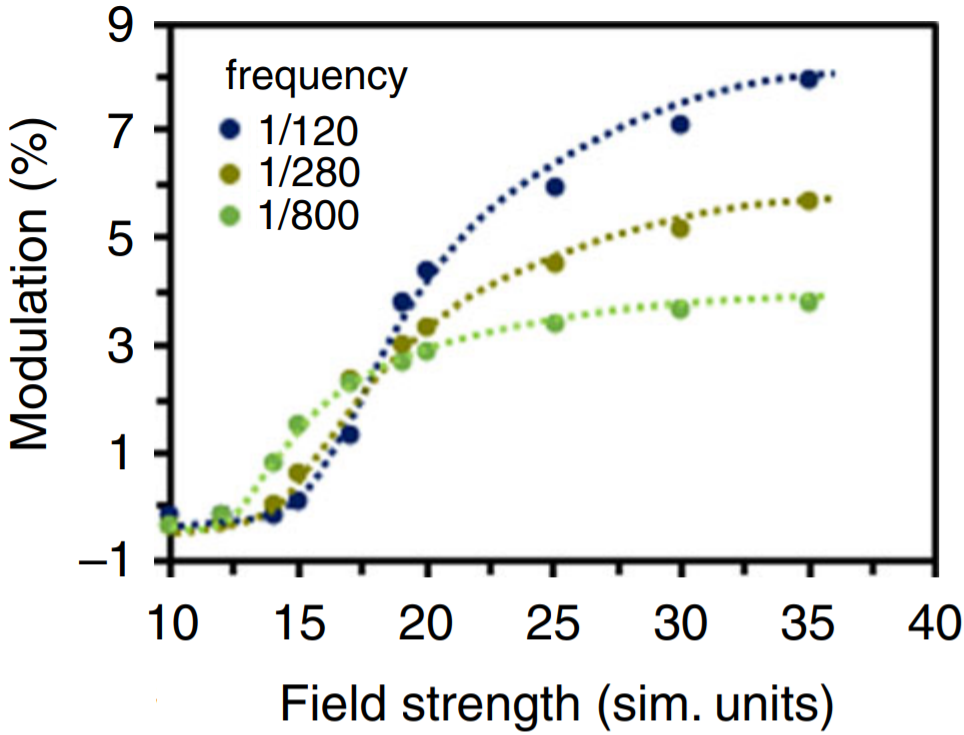}} \\
    \subfloat{\includegraphics[width=4.cm]{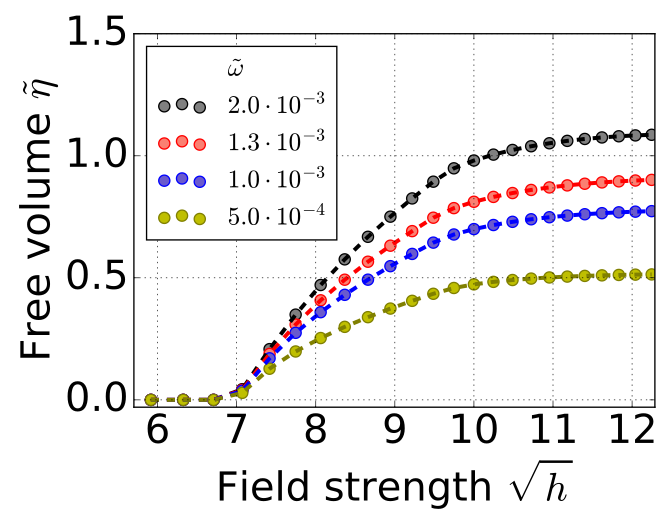}}
    \subfloat{\includegraphics[width=4.cm]{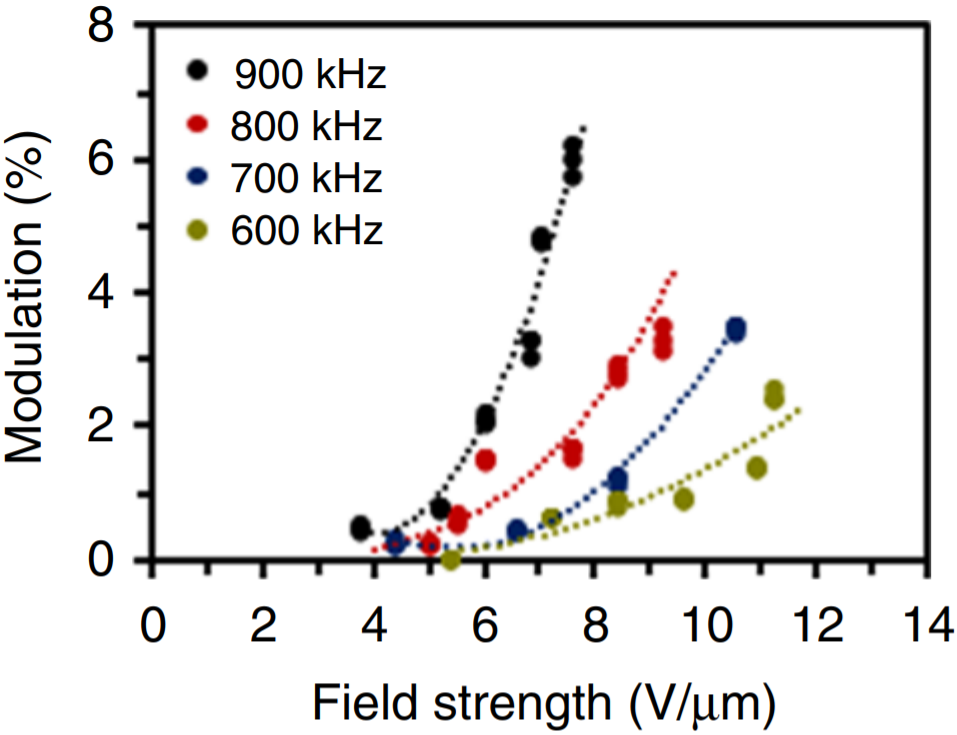}}
    \caption{Steady-state free volume gain as a function of electric field strength and driving frequency for theory (left), MD simulations (top right) and experiments (bottom right). Theory: we denote the scaled free volume order parameter $\Tilde{\eta}$, the scaled driving frequency $\Tilde{\omega}=\omega\left[\xi^2/\Gamma_{\eta}B_0^2C\right]$ and the scaled electric field strength $\sqrt{h}\propto\lvert\underline{E}\rvert$. See the Appendix for the full scaling procedure and see Figure \ref{fig:time traces} for the used parameter values. Simulations are in terms of fundamental Lennard-Jones units, at fixed temperature $T=0.35$ and fixed pressure $p=1.0$, with time steps $\diff t=0.001$, total simulation time $t_\text{sim}=10.0$ and the driving frequency $f$ in terms of $\left(100\times\diff t\right)^{-1}$; see the end of section \ref{sec:theory} and Ref. \cite{liu2017protruding} for further details. Right panels reprinted from Ref. \cite{liu2017protruding} under a Creative Commons Attribution 4.0 International License \url{http://creativecommons.org/licenses/by/4.0/}.}\label{fig:comparison}
\end{figure}

To more strongly link these different mechanisms to experimentally verifiable predictions, we subsequently perform a simultaneous sweep over both electric field strength and driving frequency, as Figure \ref{fig:comparison} shows. Focusing first on the top-left panel, which shows a range of driving frequencies encompassing the resonances for both mechanisms, we again recover two regimes. At low field strengths low-frequency actuation yields the greatest free volume generation, whereas at high field strengths high-frequency actuation yields optimal results. We remark that Liu \textit{et al.} have performed similar sweeps in their MD simulations, shown in the top-right panel \cite{liu2017protruding}, which show the same separation into two distinct regimes as a function of field strength. This suggests that the competition between two different mechanisms for free volume generation also plays an important role in the MD simulations.

The bottom-left panel in Figure \ref{fig:comparison}, alternatively, shows the theoretical prediction for the same sweep but over a narrower range of driving frequencies. As for a sufficiently narrow range of driving frequencies one effectively probes only a single mechanism, the result shows a single, monotonic regime. The trend we observe here is in line with experimental results reported by Liu \textit{et al.}, shown in the bottom-right panel, suggesting that the experimentally used range of driving frequencies is too narrow to observe the different mechanisms for free volume generation that we predict. 

Thus, from an experimental point of view the model predicts, on the one hand, that broadening the range of driving frequencies gives rise to non-monotonic behavior as a function of the driving frequency. This leads to two distinct regimes dominated by different physics, namely reorientation and complete relaxation of the dipolar mesogens at low electric field strengths and ``pumping" of the dipolar mesogens at high electric field strengths. On the other hand, the model, in accordance with MD simulations, suggests that increasing the electric field strength leads to saturation of the free volume generation at a given driving frequency.

\section{Conclusion and Outlook}\label{sec:conclusion}
In summary, we have constructed a Landau-type theory based on the principle of excluded-volume-driven free volume generation, which shows strong qualitative agreement with our own but also previous MD simulations. Subsequently extending the model to include dynamics indicates that there is a plasticization time associated with macroscopic volume gain; a finding reinforced by experiments \cite{van2019morphing}. Based on the model, we predict that this plasticization time scales inversely with the liquid crystalline mesogen aspect ratio, the cross-linking fraction and the initial orientational order cross-linked into the network. Finally, we show that the AC actuation of the model LCN is governed by two competing mechanisms, which operate at different frequencies and dominate at different electric field strengths. 

We use our Landau-theoretical framework to rationalise the observation of two distinct regimes in the MD simulation data reported by Liu \textit{et al.}, and postulate that the absence of a second regime in their experiments is due to a limited range of driving frequencies \cite{liu2017protruding}. Accordingly, we propose follow-up experiments in which the liquid crystalline mesogen aspect ratio, cross-linking fraction and the initial orientational order cross-linked into the network are varied, and the range of driving frequencies and field strengths widened. This would not only allow for a verification of the theory, but also potentially provide concrete experimental handles to optimize the design of LCNs in industrial applications, paving the way toward LCN-based devices in the fields of haptic feedback, self-cleaning surfaces and pattern formation.

%
%

\begin{acknowledgments}
	This research received funding from the Dutch Research Council (NWO) in the framework of the ENW PPP Fund for the top sectors and from the Ministry of Economic Affairs in the framework of the `PPS-Toeslagregeling'.
\end{acknowledgments}

\appendix*
\section{Scaling procedure\label{appendix}}
Throughout this paper we use the following scaling procedure. Although the scalings used for the static model are also given in the main text, here we briefly reiterate these for ease of reference. Firstly, we introduce the substitutions
\begin{equation}\label{eq:dimensionlessdynamics}
    \begin{cases}
        s_i=S_i/S_+, \, \, \, \, \, \, \, \, \, \, \, \, \, \, i=0,1,2,3\\
        a=A/A_+,\\
        h=H/\overline{H},\\
        \tilde{\eta}=\eta/\left(\xi B^2/B_0C^2\right),
    \end{cases}
\end{equation}
where $S_+=B/2C$ and $A_+=B^2/4C$ denote the spinodal of the nematic liquid crystal, and $\overline{H}=B^3/27C^2$ the critical field strength. In addition, we scale the parameters $\tilde{\kappa}\equiv\kappa/\left(B^2/C\right)$, $\zeta\equiv\xi^2/B_0C$ and $\tilde{\lambda}\equiv\lambda/\left(B^2/C\right)$, which represent the magnitude of the effective coupling to the polymer network, the coupling to the free volume and the excluded-volume-like coupling terms. Finally, we scale the parameter $\tilde{B_1}\equiv B_1/\left(B^4/C^3\right)$, associated with the bulk-modulus-like term in $x^2$. As noted in the main text, we choose the phenomenological constant $a=2s_0-s_0^2$ such that the initial configuration $s_1=s_2=s_3=s_0$, $x^2=0$ and $\tilde{\eta}=0$ coincides with the free energy minimum in the absence of an electric field, $h=0$. This eliminates $a$ as a parameter in our model in favor of $s_0$, and leaves us, for a given field strength $h$, with the set of parameters $\tilde{\kappa},\tilde{\lambda},\zeta,\tilde{B}_1$ and $s_0$ characterizing the equilibrium properties of the model. The free energy density $g$ of equation \eqref{eq:gpop} accordingly becomes $g=\tilde{g}\left(\tilde{\kappa},\tilde{\lambda},\zeta,\tilde{B}_1,s_0\right)B^4/C^3$, with $\tilde{g}$ a dimensionless free energy density.

If we subsequently consider the mean-field dynamics as a function of the time $\tau$, described by equation \eqref{eq:dynamicaleqs}, reprinted here for ease of reference
\begin{equation}\label{eq:dynamicaleqsapp}
    \partial_\tau\psi=-\Gamma_\psi\left[\frac{\partial g}{\partial \psi}+\gamma\eta\langle\tau\rangle_\eta\delta_{\psi,\eta}\right],
\end{equation}
with $\langle\tau\rangle_\eta$ the ensemble-averaged time the free volume has existed for and $\delta_{\psi,\eta}$ the Kronecker delta, we introduce a new set of parameters: the kinetic coefficients $\Gamma_\psi$, with $\psi=S_1,S_2,S_3,x,\eta$, and the phenomenological constant $\gamma$, which represents the viscoelastic relaxation of the LCN. Starting with the kinetic coefficients $\Gamma_\psi$, we can choose a convenient scaling by recognizing that all order parameter scalings are of the form $\Tilde{\psi}=\psi/\psi_\text{ref}$, where $\psi_\text{ref}=S_+,S_+,S_+,1,\xi B^2/B_0C^2$ indicates the relevant scaling. Writing the non-dimensionalized free energy density $\Tilde{g}=g \, C^3/B^4$ and scaling the time $\Tilde{\tau}=\tau/\tau_0$ with some as of yet unspecified reference time $\tau_0$, then allows us to recast the set of dynamical equations \eqref{eq:dynamicaleqsapp} in the dimensionless from
\begin{equation}\label{eq:dynamicrescale}
    \partial_{\Tilde{\tau}} \Tilde{\psi}=-\frac{\tau_0\Gamma_\psi }{\psi_\text{ref}^2}\frac{B^4}{C^3}\left[\frac{\partial\Tilde{g}}{\partial\Tilde{\psi}}+\left(\gamma \, \eta_\text{ref}^2\tau_0\frac{C^3}{B^4}\right)\tilde{\eta}\langle\tilde{\tau}\rangle_{\tilde{\eta}} \delta_{\tilde{\psi},\tilde{\eta}}\right].
\end{equation}
From equation \eqref{eq:dynamicrescale}, we read off the \textit{effective} kinetic coefficients $\Tilde{\Gamma}_{\Tilde{\psi}}=\tau_0\Gamma_\psi B^4/\psi_\text{ref}^2C^3$ that apply to the scaled order parameters $\Tilde{\psi}$. This suggests a sensible choice for the reference time is to set $\tau_0=\xi^2/\Gamma_\eta B_0^2C$, such that we have $\Tilde{\Gamma}_{\Tilde{\eta}}=1$, i.e., we measure time relative to the dynamics of the scaled free volume order parameter $\tilde{\eta}$. This leaves us with the set of scaled kinetic coefficients
\begin{equation}\label{eq:renormalised kinetic coefficients}
    \begin{cases}
        \Tilde{\Gamma}_{s_1}=\left(\Gamma_{S_1}/\Gamma_\eta\right)/\left(C^2B_0^2/4B^2\xi^2\right),\\
        \Tilde{\Gamma}_{s_2}=\left(\Gamma_{S_2}/\Gamma_\eta\right)/\left(C^2B_0^2/4B^2\xi^2\right),\\
        \Tilde{\Gamma}_{s_3}=\left(\Gamma_{S_3}/\Gamma_\eta\right)/\left(C^2B_0^2/4B^2\xi^2\right),\\
        \Tilde{\Gamma}_{x}=\left(\Gamma_{x}/\Gamma_\eta\right)/\left(C^4B_0^2/B^4\xi^2\right),\\
        \Tilde{\Gamma}_{\Tilde{\eta}}=1.
    \end{cases}
\end{equation}
We remark that this time scale is identically used to scale the driving frequency $\tilde{\omega}=\omega\tau_0$ for AC actuation, and allows us to also evaluate the scaled phenomenological constant $\tilde{\gamma}=\gamma/\left(\Gamma_\eta B_0^4C/\xi^4\right)$, which completes the scaling procedure.

\bibliography{DLT_LCN}

\end{document}